\documentclass[a4paper,ngerman]{llncs}
\pdfoutput=1
\usepackage[T1]{fontenc}
\usepackage[latin9]{inputenc}
\usepackage{babel}
\usepackage{url}
\usepackage{graphicx}
\usepackage[unicode=true,
 bookmarks=true,bookmarksnumbered=false,bookmarksopen=false,
 breaklinks=true,pdfborder={0 0 0},pdfborderstyle={},backref=false,colorlinks=false]
 {hyperref}
\hypersetup{pdftitle={Das Internet-Adressbuch bedroht unsere Privatsphäre},
 pdfauthor={Dominik Herrmann}}

\makeatletter

\pdfpageheight\paperheight
\pdfpagewidth\paperwidth

\@ifundefined{date}{}{\date{}}
\usepackage[protrusion=true,expansion=true]{microtype}

\makeatother

\begin{document}

\title{Das Internet-Adressbuch bedroht\\
unsere Privatsphäre}

\author{Dominik Herrmann}

\institute{Universität Hamburg, Fachbereich Informatik, Sicherheit in verteilten
Systemen\\
herrmann@informatik.uni-hamburg.de}
\maketitle
\begin{abstract}
Dieser Beitrag fasst ausgewählte Ergebnisse der Dissertation \glqq Beobachtungsmöglichkeiten im Domain Name System: Angriffe auf die Privatsphäre und Techniken zum Selbstdatenschutz\grqq\ \cite{Herrmann-diss} zusammen. Die Dissertation liefert neue Antworten auf die Fragen \glqq Wer kann uns im Internet überwachen?\grqq\ und \glqq Wie schützen wir uns davor?\grqq. Die Arbeit befasst sich mit dem Domain Name System (DNS), dem Adressbuch des Internets. Es wird gezeigt, dass es im DNS bislang vernachlässigte Überwachungsmöglichkeiten gibt. Insbesondere wird ein Verfahren zum verhaltensbasierten Tracking vorgestellt, mit dem die Aktivitäten von Internetnutzern unbemerkt über längere Zeiträume verfolgt werden können. Einerseits wird dadurch die Privatsphäre vieler Internetnutzer bedroht, andererseits könnten daraus neue Werkzeuge für die Strafverfolgung entstehen. Weiterhin werden neue Datenschutz-Techniken vorgeschlagen, die sicherer und benutzerfreundlicher sind als die bisherigen Ansätze.
\end{abstract}

\section{Einleitung}

\begin{figure}[t]
\begin{centering}
\includegraphics[width=0.4\textwidth]{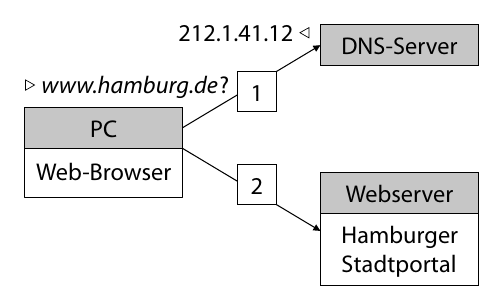}
\par\end{centering}
\caption{\label{fig:Namensaufl}Namensauflösung vor dem Aufruf einer Webseite}
\end{figure}

Nirgendwo ist unsere Privatsphäre so verletzlich wie im Internet.
Das Domain Name System (DNS) ist ein Internetdienst, den man sich
wie ein großes Adressbuch vorstellen kann. Jedes Mal, wenn man eine
Internetseite besucht, schlägt der Browser den Domainnamen der Seite
(z.\,B. \emph{www.hamburg.de}) im DNS nach. Er benötigt nämlich die
IP-Adresse (derzeit 212.1.41.12), um die Internetseite vom jeweiligen
Webserver abzurufen. Dazu kontaktiert der Browser spezielle DNS"=Server,
sog. \emph{rekursive Nameserver} (s. Abbildung~\ref{fig:Namensaufl}).
Üblicherweise stellt jeder Internetanbieter (z.\,B. T-Online) seinen
Kunden einen solchen DNS-Server zur Verfügung. Inwiefern stellt das
eine Bedrohung für die Privatsphäre dar? Der Internetanbieter kann
den Datenverkehr seiner Kunden doch ohnehin vollständig überwachen;
dass er die DNS-Anfragen seiner Kunden sehen kann, ist also kein Problem.
So wurde bisher argumentiert und genau deswegen wurden die Beobachtungsmöglichkeiten
auf den DNS-Servern bislang weitgehend vernachlässigt. 

\section{Aktuelle Entwicklungen im Domain Name System}

Die mangelnde Vertraulichkeit der Namensauflösung gewinnt allerdings
erheblich an Bedeutung. Dies lässt sich an drei aktuellen Entwicklungen
festmachen, die im Folgenden erläutert werden.

Zum einen zeichnet sich ein Trend zur \textbf{Zentralisierung der
Namensauflösung} ab. Seit etwa fünf Jahren bieten auch internationale
Konzerne wie Google, OpenDNS und Symantec kostenlose DNS"=Server
an. Diese Angebote trumpfen mit überzeugenden Vorzügen auf, etwa geringer
Verzögerung und hoher Ausfallsicherheit. Zudem haben sie bereits zur
Erhaltung der Meinungsfreiheit beigetragen. So hat die türkische Regierung
im Frühling 2014 zwar den Kommunikationsdienst Twitter mittels einer
DNS-Sperre unzugänglich gemacht. Davon ließen sich die Bürger jedoch
nicht beirren. Sie hebelten die Sperre aus, indem sie auf den DNS-Server
von Google umstiegen, der über die leicht zu merkende IP-Adresse 8.8.8.8
erreichbar ist. Später gab die Regierung nach und hob die Blockade
wieder auf.

Im Jahr 2016 beantworteten allein die DNS-Server von Google schon
mehr als 13\,\% aller DNS-Anfragen pro Tag \cite{apnic2016}. Auch
in Deutschland erfreuen sie sich zunehmender Beliebtheit. Sie werden
immer wieder von der Presse empfohlen, etwa von Spiegel Online oder
der Fachzeitschrift c\textquoteright t.

Angesichts dessen stellt sich die Frage, welche Informationen die
neuen DNS"=Anbieter über ihre Nutzer gewinnen können, inwiefern die
Nutzer solcher Angebote Nachteile befürchten müssen, und wie man sich
als Nutzer vor neugierigen Blicken des Betreibers schützen kann. Diese
Fragen werden in der Dissertation beantwortet. Die gewonnen Erkenntnisse
sind zum einen also zunächst einmal ganz unmittelbar für diejenigen
Nutzer von Bedeutung, die nicht mehr den DNS-Server ihres Internetanbieters
verwenden. Darüber hinaus profitieren Forscher und Entwickler von
den Ergebnissen der Dissertation, da sie dadurch benutzerfreundlichere
und wirksamere datenschutzfreundliche Lösungen zum Zugriff auf DNS-Server
entwickeln können.

Zweitens sind wir einer zunehmenden \textbf{Überwachung unseres Verhaltens
durch die Werbewirtschaft} ausgesetzt. Praktisch alle großen Webseiten
finanzieren sich durch Werbeanzeigen, die von sogenannten Werbenetzen
vermarktet werden. Werbenetze sind darauf spezialisiert, jedem Besucher
genau die Anzeigen zu präsentieren, die zu seinen Interessen passen.
Bisher werden die Nutzer dazu mit einer eindeutigen Tracking"=Nummer
markiert, die in einem \quotedblbase Cookie\textquotedblleft{} im
Browser abgespeichert wird. Inzwischen ziehen einige Werbenetze zur
Wiedererkennung auch weitere Merkmale heran, anhand derer sich ein
Rechner einzigartig identifizieren lässt, etwa die Liste der installierten
Schriftarten \cite{FifieldE15}. Beide Techniken erlauben es, einen
Nutzer auf seinen Streifzügen durch das Internet zu verfolgen und
ein entsprechendes Interessenprofil anzulegen. Allerdings funktioniert
diese Form der Überwachung schon heute nicht mehr zuverlässig. Den
Browser"=Herstellern ist nämlich inzwischen daran gelegen, die Privatsphäre
ihrer Nutzer zu schützen. So nehmen einige Browser, etwa der Safari"=Browser,
schon heute keine Tracking"=Cookies mehr an. Werbetreibende reagieren
auf diese Entwicklung, indem sie auf andere Tracking"=Techniken ausweichen. 

Die Dissertation zeigt auf, wie sich Nutzer durch die Auswertung ihrer
DNS"=Anfragen wiedererkennen lassen. Werbenetze könnten in Zukunft
also DNS-basierte Tracking-Techniken einsetzen, um die Aktivitäten
von Nutzern zu verfolgen. Insbesondere das Unternehmen Google ist
hier in einer guten Ausgangsposition, da es nicht nur einen öffentlichen
DNS"=Server anbietet, sondern auch das derzeit größte Werbenetz \quotedblbase Doubleclick\textquotedblleft{}
betreibt.

Hervorzuheben sind hierbei zwei Aspekte. Erstens kann grundsätzlich
\emph{jedes} Werbenetz die verhaltensbasierte Tracking-Technik einsetzen,
um Nutzer anhand ihres Surfverhaltens zu verfolgen. Zweitens ist verhaltensbasiertes
Tracking für Außenstehende überhaupt nicht mehr erkennbar. Dadurch
stellt es eine erhebliche Verletzung der informationellen Selbstbestimmung
dar.

Die Ergebnisse der Dissertation betreffen also nicht nur die Nutzer
neuer DNS-Anbieter, sondern uns alle. Sie machen deutlich, wie schwer
es schon heute ist, sich im Internet mit technischen Mitteln vor unerwünschter
Beobachtung zu schützen. Erschwerend kommt hinzu, dass alle staatlichen
Bestrebungen, die privatwirtschaftliche Überwachung zu regulieren
(vgl. die sog. Cookie-Richtlinie der Europäischen Union) und insbesondere
die Sanktionierung von Verstößen gegen die geltenden Datenschutzgesetze
natürlich ins Leere laufen, wenn der Einsatz einer Tracking-Technik
überhaupt nicht nachweisbar ist.

Drittens lässt sich ein zunehmendes Interesse an der \textbf{Auswertung
von Verkehrsdaten} (\quotedblbase Metadaten\textquotedblleft ) erkennen
\textendash{} nicht nur bei Nachrichtendiensten, sondern auch bei
polizeilichen Ermittlungsbehörden. DNS"=Server stellen einen bislang
unterschätzten, jedoch vielversprechenden Beobachtungspunkt dar. Im
Gegensatz zur Aufzeichnung und Auswertung des gesamten Datenverkehrs,
was der sprichwörtlichen Suche nach der Nadel im Heuhaufen gleicht,
stellt die Aufzeichnung und Auswertung der DNS"=Anfragen wesentlich
geringere Ressourcenanforderungen. Der Anteil des DNS-Datenverkehrs
am übertragenen Gesamtvolumen beträgt nur etwa 0,05\,\%. Aus den
aufgezeichneten DNS"=Anfragen gehen genau die Informationen hervor,
die für forensischen Untersuchungen von Interesse sind: die IP"=Adresse
eines Nutzers, der angefragte Domainname und der Zeitpunkt der Anfrage.
Darüber hinaus ließen sich durch Beschlagnahme von DNS"=Protokollen,
die von den Anbietern üblicherweise für einige Zeit zur Störungserkennung
aufbewahrt werden, auch \emph{nachträglich} noch Informationen über
die Internetaktivitäten eines Verdächtigen gewinnen. Zudem könnten
sich auch Datenschützer mit der forensischen Auswertung der DNS-Anfragen
arrangieren. Anstelle über die Einrichtung zusätzlicher Überwachungsmaßnahmen,
etwa einer allgemeinen Vorratsdatenspeicherung, nachzudenken, wäre
es besser, erst einmal die bereits existierenden Informationsquellen
auszuschöpfen. 

Allerdings besteht bei der Auswertung von DNS-Anfragen ein gewisses
Risiko von Fehlinterpretationen. Die Tatsache, dass ein Nutzer einen
Domainnamen aufgelöst hat, muss nämlich nicht zwangsläufig bedeuten,
dass er auch zugehörige Webseite abgerufen hat. Bei unsachgemäßer
Interpretation könnte es also zur Verfolgung unschuldiger Nutzer kommen.
In der Dissertation wird daher nicht nur aufgezeigt, wie bei der Auswertung
vorzugehen ist, sondern auch wie aussagekräftig die dabei gewonnen
Informationen sind.

Zusammengefasst besteht die Bedeutung der vorliegenden Dissertation
einerseits darin, auf neue Beobachtungsmöglichkeiten aufmerksam zu
machen, die unsere Privatsphäre im Internet bedrohen, und zum anderen
darin, gestaltungsorientiert Techniken zum Schutz vor unerwünschter
Überwachung vorzuschlagen.

\section{Was können DNS-Server über ihre Nutzer herausfinden?}

Zunächst wird in der Dissertation der Frage nachgegangen, inwiefern
ein DNS-Anbieter anhand der aufgelösten Domains nachvollziehen kann,
welche Webseiten ein Nutzer abruft. Das ist nämlich gar nicht ohne
weiteres möglich.

Der Anbieter steht dabei vor zwei Herausforderungen. Erstens kommt
beim DNS"=Server \emph{nur} der Domainname an, also nicht die vollständige
Adresse der besuchten Seite. Bei bestimmten Webseiten, etwa \emph{\url{http://de.wikipedia.org/wiki/Alkoholkrankheit}},
ist allerdings gerade die URL problematisch, wohingegen die Domain
an sich (\emph{de.wikipedia.org}) vergleichsweise unverfänglich ist.
Zweitens stimmen die abgerufenen Webseiten nicht mit den beobachtbaren
Domains überein, d.\,h. aus der Beobachtung einer DNS-Anfrage für
eine bestimmte Domain (z.\,B. \emph{\url{www.anonyme-alkoholiker.de}})
kann der Anbieter nicht unbedingt schlussfolgern, dass ein Nutzer
auch die zugehörige Webseite besucht hat. Beim Abruf einer Webseite
stellt ein Web-Browser nämlich häufig mehr als eine DNS"=Anfrage,
typischerweise werden zwischen zehn und 20 Domains aufgelöst. Diese
zwei Probleme erschweren die Bestimmung der besuchten Webseiten erheblich.

Mit dem in der Dissertation entwickelten \textbf{Website"=Fingerprinting"=Verfahren}
lassen sich diese Probleme allerdings überwinden. Die durchgeführten
Analysen zeigen, dass viele Webseiten ein so charakteristisches DNS-Abrufmuster
erzeugen, dass sich ihr Abruf daran mehr oder weniger zweifelsfrei
erkennen lässt. Der DNS"=Anbieter könnte eine Datenbank anlegen,
in der die Abrufmuster aller für ihn interessanten Webseiten enthalten
sind. Um die von einem Nutzer besuchten Webseiten zu bestimmen, müsste
der Anbieter die Anfragen des Nutzers dann lediglich mit den Abrufmustern
in seiner Datenbank abgleichen. Welche Erfolgsaussichten dieser Ansatz
hat, wurde in der Dissertation genauer untersucht.

So stellte sich heraus, dass Reihenfolge und Zeitabstände zwischen
den einzelnen DNS"=Anfragen außen vor bleiben sollten, um die Robustheit
der Erkennung zu erhöhen. Das Abrufmuster der Webseite \emph{www.margersucht.de}
ist beispielsweise die Menge \{\emph{www.amazon.de}, \emph{www.ess\-frust.de},
\emph{www.essstoerungen-frankfurt.de}, \emph{www.magersucht.de}, \emph{www.telefonseelsorge.de}\};
das Abrufmuster des o.\,g. Wikipedia-Eintrags zu \quotedblbase Alkoholkrankheit\textquotedblleft{}
enthält mehr als 30~Domains (u.\,a. \emph{de.wikipedia.org}, \emph{bits.wikimedia.org},
\emph{counsellingresource.com}, \emph{www.spiegel.de}, \emph{www.stadt-und-gemeinde.de}
und \emph{w210.ub.uni-tuebingen.de}).

Dieser Ansatz ist durchaus vielversprechend. Von den 5000 beliebtesten
Wikipedia"=Einträgen hatten im Versuch 98,9\,\% ein einzigartiges
Abrufmuster \textendash{} ihr Abruf wäre für den DNS"=Server also
sehr wahrscheinlich erkennbar. Auch in weiteren Untersuchungen mit
5000 zufällig ausgewählten Wikipedia"=Einträgen, 6200 News"=Beiträgen
von \emph{www.heise.de} und den Homepages von 100\,000 populären
Webseiten waren bei der überwiegenden Mehrheit der Seiten einzigartige
Abrufmuster zu beobachten.

Aber wieso ist es ein Problem, dass der DNS"=Server herausfinden
kann, welche Webseiten seine Nutzer abrufen? Weil Nutzern daraus erhebliche
Nachteile entstehen können, die sie gar nicht bemerken. Anhand der
besuchten Webseiten lassen sich mitunter Geschlecht, Alter, Beruf
und die aktuellen Interessen eines Nutzers erschließen. ,,Diese Rückschlüsse
können zutreffend sein oder vollkommen abwegig``, erläutert Peter
Schaar, ehemaliger Bundesdatenschutzbeauftragter, in einem Interview
auf \emph{welt.de,} ,,aber diese Informationen wären mit hoher Wahrscheinlichkeit
für den Anbieter einer Berufsunfähigkeitsversicherung, die Sie vielleicht
gerne abschließen würden, von großem Interesse. Der könnte Ihnen deswegen
vielleicht einen Vertrag verweigern oder höhere Beiträge verlangen.``

Abgesehen davon kann ein DNS"=Anbieter auch auf die \textbf{von einem
Nutzer eingesetzten Anwendungen} schließen. Viele Anwendungen verraten
sich bei der Suche nach Updates (z.\,B. \emph{windowsupdate.com},
\emph{su3.mcaffee.com}, \emph{aus3.mozilla.org}). Auf diese Anfragen
ist der DNS"=Anbieter jedoch nicht unbedingt angewiesen: Da der genaue
Ablauf der Namensauflösung nicht präzise festgelegt ist, haben sich
unterschiedliche Varianten etabliert. Charakteristisch ist z.\,B.
die Zeitspanne, die ein System auf einen DNS"=Server höchstens wartet,
bevor es seine DNS"=Anfrage erneut übermittelt. Die in der Dissertation
vorgestellten Untersuchungen zeigen, dass sich gängige Betriebssysteme
und Web"=Browser daran gut unterscheiden lassen.

Anhand der DNS"=Anfragen lassen sich also detaillierte Informationen
über die Betriebsumgebung eines Nutzern gewinnen. Dies ist insofern
beunruhigend, da dieses Wissen die Vorbereitung gezielter Angriffe
(sog. ,,targeted attacks``) erleichtert, bei denen dem Opfer Schadsoftware
untergeschoben wird. Wenn diese Schadsoftware genau auf die vorhandene
Infrastruktur des Opfers abgestimmt ist, steigt die die Wahrscheinlichkeit,
dass der Angriff erfolgreich ist. Sie kann dann nämlich Sicherheitsmechanismen
wie Virenscanner oder Firewalls umgehen.

\section{Lassen sich Nutzer anhand ihres Verhaltens wiedererkennen?}

Wenn ein DNS"=Server lediglich beobachten könnte, dass man gerade
dabei ist, eine bestimmte Webseite abzurufen, wäre das Risiko für
die eigene Privatsphäre relativ überschaubar. Besonders aussagekräftige
Interessen- und Nutzungsprofile entstehen allerdings, wenn ein DNS"=Anbieter
die Aktivitäten eines Nutzers über Tage hinweg verfolgen kann. Dann
treten nämlich wiederkehrende Muster zu Tage, anhand derer der Anbieter
auf das soziale Umfeld, Gewohnheiten und Krankheiten eines Nutzers
schließen könnte \textendash{} genau deswegen geht vielen Menschen
das eingangs erwähnte Tracking durch Werbenetze zu weit.

Zum Glück ist eine solch langfristige Überwachung bei den meisten
Nutzern aber gar nicht möglich. Viele Internetprovider vergeben schließlich
sogenannte \quotedblbase dynamische IP-Adressen\textquotedblleft ,
d.\,h. ihre Kunden sind jeden Tag unter einer anderen IP-Adresse
online. Aus Sicht des Datenschutzes sind dynamische IP-Adressen sehr
zu begrüßen. Der DNS"=Server verliert diese Nutzer bei jedem Wechsel
der IP-Adresse nämlich aus den Augen. Gleiches gilt auch für Werbenetze,
wenn man die im Browser die Cookies löscht (und der Browser keinen
einzigartigen Fingerabdruck hat). Zumindest dachte man das bisher.

Die Dissertation zeigt, dass diese Einschätzung wohl revidiert werden
muss. In der Dissertation wurde ein Verfahren entwickelt, mit dem
ein DNS"=Anbieter oder ein Werbenetz einen Nutzer trotz wechselnder
IP"=Adresse (und ggf. gelöschter Cookies) wiedererkennen kann. Das
Verfahren basiert auf der Annahme, dass jeder Nutzer beim Surfen seinen
Interessen und Vorlieben nachgeht, und dass jeder Nutzer eine einzigartige
Kombination von Interessen und Vorlieben hat.

Das \textbf{verhaltensbasierte Verkettungsverfahren} funktioniert
wie folgt: Der DNS-Anbieter sammelt fortlaufend die Domains aller
Webseiten, die alle Nutzer (identifiziert durch ihre jeweilige IP-Adresse)
innerhalb eines Tages auflösen. Die besuchten Webseiten und die Anzahl
der Anfragen, die auf jede Webseite entfallen, werden abgespeichert.
Sie stellen den Fingerabdruck eines Nutzers dar, also das, was auf
unseren Fingerkuppen Schleifen, Wirbel und Minuzien sind. Ein Auszug
aus einem solchen Fingerabdruck könnte etwa lauten: \quotedblbase Nutzer
\quotesinglbase 371241\textquoteright{} war am 01.03.2015 insgesamt
12 Mal auf www.google.com, neun Mal auf www.hamburg.de, zwei Mal auf
intranet.lufthansa.com und ein Mal auf www.bundestag.de\textquotedblleft .

\begin{figure}[t]
\begin{centering}
\includegraphics[width=0.4\textwidth]{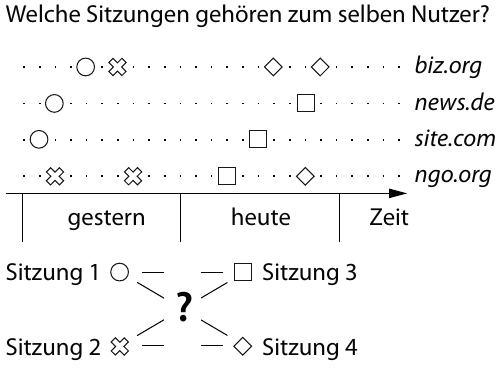}
\par\end{centering}
\caption{\label{fig:Verkettung-von-Sitzungen}Verkettung von Sitzungen anhand
der besuchten Webseiten}
\end{figure}

Nehmen wir an, dass Nutzer 371241 in seiner nächsten Internetsitzung
am 02.03.2015 unter eine andere IP-Adresse hat, er zwischenzeitlich
alle Tracking-Cookies gelöscht hat, und dass sein Browser keinen einzigartigen
Fingerabdruck besitzt. Eigentlich sollten DNS-Server oder Werbenetze
in diesem Fall keine Verbindung zwischen der aktuellen Sitzung und
der vorherigen Sitzung von Nutzer 371241 herstellen können (s. Abbildung~\ref{fig:Verkettung-von-Sitzungen}).
Durch verhaltensbasierte Verkettung könnte ihnen das allerdings trotzdem
gelingen. Dazu müssen sie lediglich die Webseiten und Anfragehäufigkeiten
aus der aktuellen Sitzung des Nutzers mit allen Fingerabdrücken in
ihrer Datenbank vergleichen. Für den Vergleich bieten sich bewährte
Algorithmen aus dem Data-Mining an. In der Dissertation kommt u.\,a.
ein Naive-Bayes-Klassifikator \cite{ManningRaghavanSchuetze08} zum
Einsatz, der häufig in Spam-Filtern eingesetzt wird, um vollautomatisch
E-Mails mit unerwünschter Werbung auszusortieren.

Damit der Klassifikator Nutzer 371241 anhand seines Verhaltens wiedererkennen
kann, müssen sich dessen Sitzungen zum einen von den Sitzungen der
anderen Nutzer ausreichend stark unterscheiden. Das allein reicht
jedoch nicht. Die eigentlich interessante Frage ist, ob wir unsere
Interessen und Vorlieben so regelmäßig im Internet ausleben, dass
wir daran jeden Tag wiedererkannt werden können.

Diese Fragestellung wurde in einer umfangreichen Studie untersucht.
Dazu wurden in Kooperation mit dem Rechenzentrum der Universität Regensburg
die DNS"=Anfragen von Angestellten und Studierenden über einen Zeitraum
von fünf Monaten protokolliert. Zum Schutz der Privatsphäre wurden
die IP-Adressen der Nutzer durch statische Pseudonyme ersetzt. Da
im Campus"=Netz jedem Nutzer immer dieselbe IP"=Adresse zugewiesen
wurde, konnten die Vorhersagen des entwickelten Verkettungsverfahrens
mit der tatsächlichen Zuordnung verglichen werden.

Mit diesem Datensatz wurden zahlreiche Experimente unter realitätsnahen
Bedingungen durchgeführt. Die Wiedererkennung gelingt nicht immer,
aber überraschend häufig. In einem der Experimente wird beispielsweise
eine Konstellation mit mehr als 3800 Studenten über einen Zeitraum
von zwei Monaten betrachtet. Dabei wird ein DNS"=Anbieter simuliert,
der die verhaltensbasierte Verkettung nutzt, um die Sitzungen aller
Studenten von einem Tag auf den nächsten zu verknüpfen. Im Mittel
wäre ihm dies in 86\% der Fälle gelungen. Bei knapp 14\% der Studenten
hätte er sogar alle Sitzungen korrekt miteinander verbunden. Auch
bei größeren Nutzergruppen gelingt die Verknüpfung der Sitzungen noch
erstaunlich gut. In einem Experiment mit mehr als 12\,000 Nutzern
wurde zum Beispiel noch eine Genauigkeit von 76\% erreicht. Weitere
Details zur Methodik stehen in \cite{HBF:2013}.

Täglich wechselnde IP"=Adressen schützen demnach weitaus schlechter
vor Langzeit"=Überwachung als bislang angenommen.

Unklar bleibt allerdings wie gut die Verkettung bei \emph{normalen
}Werbenetzen funktionieren würde, also bei Werbenetzen, die (im Unterschied
zu Google) keinen eigenen DNS"=Server betreiben. Eine genaue Prognose
ist anhand des zur Verfügung stehenden DNS"=Datensatzes nicht möglich.
Normale Werbenetze sehen schließlich nur die Anfragen für Webseiten,
auf denen sie ihre Werbung vermarkten.

Die Großen der Branche sind allerdings auf nahezu jeder beliebten
Webseite vertreten \cite{EnglehardtN16} \textendash{} und das könnte
in der Praxis durchaus ausreichen, um Nutzer wiederzuerkennen: Im
Experiment sank die Genauigkeit nämlich nur um sechs Prozentpunkte,
wenn dem Klassifikator \textendash{} anstelle sämtlicher DNS"=Anfragen
\textendash{} nur die die Anfragen für die 500 beliebtesten Domains
zur Verfügung standen. Betroffen sind also nicht nur Nutzer, die besonders
seltene Webseiten besuchen. Erkennbar ist man offenbar auch dann,
wenn man nur im Mainstream schwimmt.

\section{Wie schützen wir uns vor der Überwachung durch DNS-Server?}

Da nicht davon auszugehen ist, dass das DNS auf absehbare Zeit von
einem datenschutzfreundlicheren Namensdienst abgelöst wird, bleibt
den Nutzern lediglich die Möglichkeit, Techniken zum Selbstdatenschutz
einzusetzen. Zunächst denkt man hier natürlich an die bereits existierenden
Techniken. Alle in der Praxis verfügbaren Lösungen weisen aber unangenehme
Einschränkungen auf. So taucht man zwar in der Masse unter, wenn man
einen Anonymisierer wie Tor \cite{Dingledine:2004} und einen restriktiv
konfigurierten Browser verwendet; allerdings funktionieren einige
Webseiten dann nicht mehr ordnungsgemäß und die Ladezeiten steigen
erheblich. Etwas besser schneiden speziell angepasste Anonymisierungsdienste
ab \cite{HFLF14}. Diese schützen allerdings nur vor der Beobachtung
durch die Betreiber von DNS-Servern.

Zhao und Kollegen haben vorgeschlagen, die aufzulösenden Domainnamen
durch einige zufällig gewählte Dummy"=Anfragen zu verschleiern \cite{Castillo-Perez:2008,Zhao:2007a}.
Die Dissertation zeigt, dass dieses Range"=Query"=Verfahren weit
weniger Schutz bietet als bislang angenommen \cite{HerrmannMF14}.
Bei 9~Dummys kann ein findiger DNS"=Anbieter bis zu 94\,\% der
untersuchten Seiten erkennen, bei 99 Dummys immerhin noch 91\,\%.
Wie kommt es zu dieser überraschend hohen Genauigkeit?

Der DNS"=Anbieter wendet lediglich das oben vorgestellte Website"=Fingerprinting"=Verfahren
an. Wenn die Dummy-Domains zufällig ausgewählt werden, ist es relativ
unwahrscheinlich, dass dabei ein vollständiges Abrufmuster irgendeiner
anderen Webseite entsteht. Neben dem vollständig vorhandenen (tatsächlichen)
Abrufmuster treten daher lediglich zahlreiche unvollständige Abrufmuster
auf, die der Anbieter leicht ausschließen kann.

Auf Basis dieser Erkenntnis wurde in der Dissertation ein verbessertes
Range"=Query"=Verfahren vorgeschlagen, das die Dummy"=Domains so
auswählt, dass stets \textbf{vollständige Dummy-Abrufmuster} entstehen.
Es bietet zwar wirksamen Schutz vor Website"=Fingerprinting, sein
Einsatz gestaltet sich allerdings vergleichsweise aufwändig. Hier
sind weitere Untersuchungen nötig, um die Praktikabilität zu verbessern.

Einen anderen Ansatz verfolgt der in der Dissertation entwickelte
\textbf{DNSMIX"=Anonymitätsdienst} \cite{FederrathFHP11-dnsmixes}.
Hier werden nicht die beabsichtigten Domains verschleiert, sondern
die \emph{Identität der Teilnehmer}. Als Basis des Dienstes fungiert
eine Mix-Kaskade, die eine anonyme Namensauflösung ermöglicht. Die
Besonderheit des DNSMIX-Dienstes ist jedoch der sog. \textbf{Push-Dienst},
welcher die IP-Adressen aller häufig aufgelösten Domains unverlangt
an alle mit dem Dienst verbundenen Teilnehmer übermittelt \textendash{}
diese Domains können die Nutzer dann völlig autonom, also für Außenstehende
unbeobachtbar, auflösen. Dazu muss der Push-Dienst kontinuierlich
überprüfen, ob sich die IP-Adressen der betreffenden Domains geändert
haben.

Wäre ein solcher Push-Dienst überhaupt praktikabel? Ist der Aufwand
für die Aktualisierung nicht viel zu groß? Um diese Fragen zu beantworten,
wurden im Rahmen der Dissertation Untersuchungen mit dem DNS"=Daten\-verkehr
von 2000 Nutzern durchgeführt. Dabei stellte sich heraus, dass der
Großteil der DNS"=Anfragen einer relativ kleinen Menge von Domains
gilt.

\begin{figure}[t]
\begin{centering}
\includegraphics{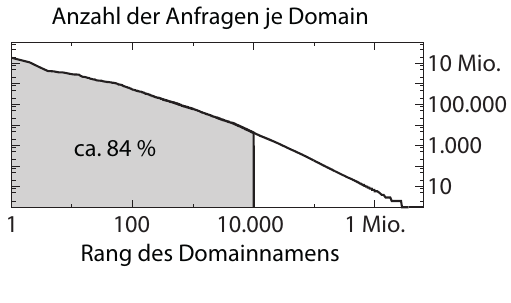}
\par\end{centering}
\caption{\label{fig:Das-Anfragevolumen-f=0000FCr}Das Anfragevolumen für eine
Domain hängt von ihrer Popularität ab.}
\end{figure}

Als guter Kompromiss zwischen Kosten und Nutzen hat sich die Fokussierung
auf die 10\,000 am häufigsten angefragten Domains erwiesen. Wenn
diese vom Push-Dienst an die Nutzer übermittelt werden, können die
Nutzer im Mittel mehr als 80 \% ihrer Anfragen einsparen (s. Abbildung~\ref{fig:Das-Anfragevolumen-f=0000FCr}).
Da sich die IP-Adressen dieser Domains nur vergleichsweise selten
ändern werden für die kontinuierliche Übermittlung nur etwa 0,8 KB/s
je Nutzer benötigt. Den Großteil der Namensauflösung können die Nutzer
somit verzögerungsfrei und für Außenstehende völlig unbeobachtbar
abwickeln. Die übrigen Anfragen lösen sie verschlüsselt über die Mix"=Kaskade
auf. Die Antwortzeiten waren dabei im Mittel mit 0,17\,s deutlich
niedriger als bei Tor.

\section{Lässt sich verhaltensbasierte Verkettung komfortabel verhindern?}

Wirksamer Schutz vor jeglicher Beobachtung durch den DNS"=Server
ist also grundsätzlich umsetzbar, allerdings immer mit Aufwand verbunden.
Die Nutzer müssten zumindest eine zusätzliche Software, z.\,B. den
DNSMIX-Client, auf ihren Rechnern installieren. Viele Nutzer sind
dazu allerdings nicht bereit.

Erfreulicherweise deuten die Ergebnisse der Dissertation darauf hin,
dass es zumindest eine benutzerfreundliche Möglichkeit gibt, die verhaltensbasierte
Verkettung von Sitzungen zu verhindern. Dazu muss die\textbf{ Sitzungsdauer}
\textbf{verkürzt werden}. Die verhaltensbasierte Verkettung kann dadurch
erheblich erschwert werden: Bei stündlichem Wechsel der IP-Adresse
waren beispielsweise nur noch 55\,\% der aufeinanderfolgenden Sitzungen
verkettbar; wird die IP-Adresse alle 5~Minuten gewechselt, sinkt
die Genauigkeit auf 31\%. Enthalten die Sitzungen also nur wenige
Aktivitäten und wechselt man konsequent in kurzen Intervallen die
eigene IP-Adresse, verliert sich die Spur relativ schnell im Sande.

Ein IP-Wechsel lässt sich schon heute bewerkstelligen, wenn man einen
Internetanbieter nutzt, der bei jeder Einwahl eine andere IP-Adresse
vergibt. Eine Neueinwahl kann man durch eine Verbindungstrennung,
also etwa durch das Ziehen des Telefonkabels, erzwingen. Besonders
benutzerfreundlich ist das in der Tat allerdings nicht.

Mit Unterstützung des Internetproviders könnte der Adresswechsel in
Zukunft allerdings automatisiert werden, also völlig ohne Zutun des
Nutzers im Hintergrund stattfinden. Gerade die derzeit laufende Einführung
von IPv6 bietet hier eine Chance, die Privatsphäre aller Internetnutzer
besser zu schützen. Möglicherweise ließe sich der schnelle Adresswechsel
auch bei den heutigen IPv4-Internetanschlüssen reibungslos durchführen
\cite{corrabs-1211-4704}. In zukünftigen Forschungsprojekten soll
untersucht werden, inwiefern dieser Ansatz praxistauglich ist. 

\section{Zusammenfassung und Ausblick}

Die Dissertation zeigt, dass Nameserver über bislang vernachlässigte
Beobachtungsmöglichkeiten verfügen, welche die Privatsphäre der Nutzer
bzw. die Sicherheit ihrer IT-Systeme bedrohen. Da Internetnutzer nicht
erkennen können, ob und für welche Zwecke ihre DNS-Anfragen ausgewertet
werden, wird dadurch ihr Recht auf informationelle Selbstbestimmung
verletzt. Die Analyseverfahren haben allerdings den Charakter einer
\textbf{Dual-Use-Technologie}, d.\,h. sie eignen sich auch für den
verantwortungsvollen Einsatz im Rahmen der IT-Forensik \cite{HerrmannFF14}.
Ermittlungsbehörden können damit zukünftig z.\,B. die Internetaktivitäten
von Verdächtigen besser nachvollziehen bzw. IT-Systeme, die bei Straftaten
verwendet wurden, anhand der installierten Software einem Verdächtigen
zuordnen. Der Mechanismus zum verhaltensbasierten Tracking wurde zwischenzeitlich
weiter verbessert \cite{Herrmann-pst,KirchlerHLK16}.

Die Ergebnisse zur verhaltensbasierten Verkettung betreffen nicht
nur Nutzer von DNS-Fremdanbietern \textendash{} sie sind von grundsätzlicher
Bedeutung für alle Internetnutzer. Da die Tracking-Cookies der Werbenetze
heute von den meisten Web-Browsern nicht mehr akzeptiert werden, suchen
Werbetreibende kontinuierlich nach neuen Techniken zur Überwachung
des Nutzungsverhaltens. Es ist daher absehbar, dass Werbenetze in
Zukunft zur Wiedererkennung von Nutzern auch das Surfverhalten der
Nutzer heranziehen werden. Dieser Schritt stellt Datenschützer vor
völlig neue Herausforderungen: Im Unterschied zu Tracking-Cookies
oder Browser"=Fingerprinting"=Techniken ist der Einsatz des verhaltensbasierten
Trackings auf den Endgeräten nämlich nicht nachweisbar, da es ausschließlich
auf passiver Beobachtung beruht.

Die Ergebnisse der Dissertation sollen Internetanbieter, Software-Entwickler
und Datenschutzbeauftragte für die Beobachtungsmöglichkeiten sensibilisieren,
und langfristig die Entwicklung performanter und benutzerfreundlicher
Techniken zur datenschutzfreundlichen Namensauflösung vorantreiben.
Das in der Dissertation vorgestellte DNSMIX-Konzept zeigt, dass sich
neben der kontinuierlichen Verbesserung generischer Anonymitätsdienste
wie Tor auch die Erforschung dienstspezifischer Datenschutz"=Techniken
lohnen kann.

Viele Nutzer legen allerdings mehr Wert auf Komfort und Performanz
als auf den Schutz ihrer Daten. Die von Datenschützern ausgesprochene
Empfehlung, zum Schutz der Privatsphäre auf potenziell problematische
Innovationen \textendash{} z.\,B. DNS-Prefetching oder DNS-Fremdanbieter
\textendash{} zu verzichten, verhallt bei diesen Nutzern ungehört.
Vielversprechend sind daher insbesondere solche Lösungen, die ohne
Komfort- und Performanzeinbußen nutzbar sind \textendash{} etwa die
Verkürzung der Sitzungsdauer, die praktisch ohne Zutun der Nutzer
eingeführt werden könnte. Hier bieten sich gute Möglichkeiten der
interdisziplinären Zusammenarbeit: Erste Gespräche mit den Landesdatenschutzbeauftragten
verliefen vielversprechend und resultierten in der expliziten Empfehlung,
dass Internetzugangsanbieter ihren Kunden stets mehrere nicht zusammenhängende
IPv6-Präfixe zur Verfügung stellen sollten.

\bibliographystyle{plain}
\bibliography{arxiv}

\begin{thebibliography}{10}

\bibitem{apnic2016}
{APNIC Labs}.
\newblock {DNSSEC Validation}, 2016.
\newblock \url{http://stats.labs.apnic.net/dnssec}, accessed: 2016-10-01.

\bibitem{Castillo-Perez:2008}
Sergio Castillo-Perez and Joaqu\'{\i}n Garc\'{\i}a-Alfaro.
\newblock {Anonymous Resolution of DNS Queries}.
\newblock In {\em Proc. On the Move to Meaningful Internet Systems (OTM 2008)},
  volume 5332 of {\em LNCS}, pages 987--1000. Springer, 2008.

\bibitem{Dingledine:2004}
Roger Dingledine, Nick Mathewson, and Paul~F. Syverson.
\newblock {Tor: The Second--Generation Onion Router}.
\newblock In {\em Proc. 13th USENIX Security Symposium}, pages 303--320.
  USENIX, 2004.

\bibitem{EnglehardtN16}
Steven Englehardt and Arvind Narayanan.
\newblock {Online Tracking: A 1-million-site Measurement and Analysis}.
\newblock In Edgar~R. Weippl, Stefan Katzenbeisser, Christopher Kruegel,
  Andrew~C. Myers, and Shai Halevi, editors, {\em Proceedings of the 2016 {ACM}
  {SIGSAC} Conference on Computer and Communications Security (CCS 2016)},
  pages 1388--1401. {ACM}, 2016.

\bibitem{FederrathFHP11-dnsmixes}
Hannes Federrath, Karl-Peter Fuchs, Dominik Herrmann, and Christopher Piosecny.
\newblock {Privacy-Preserving DNS: Analysis of Broadcast, Range Queries and
  Mix-Based Protection Methods}.
\newblock In Vijay Atluri and Claudia D\'{\i}az, editors, {\em ESORICS}, volume
  6879 of {\em LNCS}, pages 665--683. Springer, 2011.

\bibitem{FifieldE15}
David Fifield and Serge Egelman.
\newblock {Fingerprinting Web Users Through Font Metrics}.
\newblock In Rainer B{\"{o}}hme and Tatsuaki Okamoto, editors, {\em Financial
  Cryptography and Data Security ({FC} 2015)}, volume 8975 of {\em Lecture
  Notes in Computer Science}, pages 107--124. Springer, 2015.

\bibitem{Herrmann-diss}
Dominik Herrmann.
\newblock {\em {Beobachtungsm{\"o}glichkeiten im Domain Name System Angriffe
  auf die Privatsph{\"a}re und Techniken zum Selbstdatenschutz}}.
\newblock Springer Vieweg, Wiesbaden, 2016.
\newblock zugleich: Dissertation, Universit{\"a}t Hamburg, 2014.

\bibitem{corrabs-1211-4704}
Dominik Herrmann, Christine Arndt, and Hannes Federrath.
\newblock {IPv6 Prefix Alteration: An Opportunity to Improve Online Privacy}.
\newblock {\em CoRR}, abs/1211.4704, 2012.
\newblock Available at \url{http://arxiv.org/abs/1211.4704}.

\bibitem{HBF:2013}
Dominik Herrmann, Christian Banse, and Hannes Federrath.
\newblock {Behavior-based Tracking: Exploiting Characteristic Patterns in DNS
  Traffic}.
\newblock {\em Computers \& Security}, 39A:17--33, November 2013.

\bibitem{HerrmannFF14}
Dominik Herrmann, Karl{-}Peter Fuchs, and Hannes Federrath.
\newblock {Fingerprinting Techniques for Target-oriented Investigations in
  Network Forensics}.
\newblock In {\em Sicherheit}, volume 228 of {\em {LNI}}, pages 375--390. {GI},
  2014.

\bibitem{HFLF14}
Dominik Herrmann, Karl{-}Peter Fuchs, Jens Lindemann, and Hannes Federrath.
\newblock {EncDNS: A Lightweight Privacy-Preserving Name Resolution Service}.
\newblock In Miroslaw Kutylowski and Jaideep Vaidya, editors, {\em {ESORICS},
  Part {I}}, volume 8712 of {\em LNCS}, pages 37--55. Springer, 2014.

\bibitem{Herrmann-pst}
Dominik Herrmann, Matthias Kirchler, Jens Lindemann, and Marius Kloft.
\newblock {Behavior-Based Tracking of Internet Users with Semi-Supervised
  Learning}.
\newblock In {\em 14th Annual Conference on Privacy, Security and Trust (PST
  2016)}. IEEE Computer Society, 2016.

\bibitem{HerrmannMF14}
Dominik Herrmann, Max Maa{\ss}, and Hannes Federrath.
\newblock {Evaluating the Security of a DNS Query Obfuscation Scheme for
  Private Web Surfing}.
\newblock In Nora Cuppens{-}Boulahia, Fr{\'{e}}d{\'{e}}ric Cuppens, Sushil
  Jajodia, Anas Abou~El Kalam, and Thierry Sans, editors, {\em Proc. 29th
  {IFIP} {TC}-11 International Conference ({SEC} 2014)}, volume 428 of {\em
  {IFIP} AICT}, pages 205--219. Springer, 2014.

\bibitem{KirchlerHLK16}
Matthias Kirchler, Dominik Herrmann, Jens Lindemann, and Marius Kloft.
\newblock {Tracked Without a Trace: Linking Sessions of Users by Unsupervised
  Learning of Patterns in Their DNS Traffic}.
\newblock In David~Mandell Freeman, Aikaterini Mitrokotsa, and Arunesh Sinha,
  editors, {\em Proceedings of the 2016 {ACM} Workshop on Artificial
  Intelligence and Security, AISec@CCS 2016}, pages 23--34. {ACM}, 2016.

\bibitem{ManningRaghavanSchuetze08}
Christopher~D. Manning, Prabhakar Raghavan, and Hinrich Sch{\"u}tze.
\newblock {\em {Introduction to Information Retrieval}}.
\newblock Cambridge University Press, Cambridge, UK, 2008.

\bibitem{Zhao:2007a}
Fangming Zhao, Yoshiaki Hori, and Kouichi Sakurai.
\newblock {Analysis of Privacy Disclosure in DNS Query}.
\newblock In {\em Proc. International Conference on Multimedia and Ubiquitous
  Engineering (MUE 2007)}, pages 952--957. IEEE, 2007.

\end{thebibliography}

\end{document}